\documentclass[sigconf]{acmart}
\usepackage{booktabs}
\usepackage{algorithm}
\usepackage{algorithmic}
\usepackage{amsmath,amsfonts}
\usepackage{algorithmic}
\usepackage{graphicx}
\usepackage{textcomp}
\usepackage{xcolor}
\usepackage{setspace}
\usepackage{multirow}
\usepackage{subfigure}
\usepackage{xspace}
\usepackage{enumitem}
\usepackage{newfloat}
\usepackage{listings}
\AtBeginDocument{%
  \providecommand\BibTeX{{%
    \normalfont B\kern-0.5em{\scshape i\kern-0.25em b}\kern-0.8em\TeX}}}

\copyrightyear{2022} 
\acmYear{2022} 
\setcopyright{rightsretained} 

\acmConference[KDD '22]{Proceedings of the 28th ACM SIGKDD Conference on Knowledge Discovery and Data Mining}{August 14--18, 2022}{Washington, DC, USA}
\acmBooktitle{Proceedings of the 28th ACM SIGKDD Conference on Knowledge Discovery and Data Mining (KDD '22), August 14--18, 2022, Washington, DC, USA}
\acmISBN{978-1-4503-9385-0/22/08}
\acmDOI{10.1145/3534678.3539108}

\usepackage{etoolbox}
\makeatletter
\patchcmd{\maketitle}{\@copyrightpermission}{
   \begin{minipage}{0.3\columnwidth}
     \href{https://creativecommons.org/licenses/by/4.0/}{\includegraphics[width=0.90\textwidth]{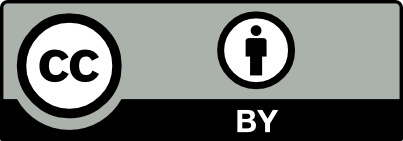}}
   \end{minipage}\hfill
   \begin{minipage}{0.7\columnwidth}
     \href{https://creativecommons.org/licenses/by/4.0/}{This work is licensed under a Creative Commons Attribution International 4.0 License.}
   \end{minipage}
  
   \vspace{5pt}
}{}{}

\makeatother

\begin{document}

\title{\our: Eliminating the User Confounding Bias for Causal Multi-touch Attribution}


\author{Di Yao}
\affiliation{
    \institution{
    Institute of Computing Technology, Chinese Academy of Sciences\\
    \country{China}
    }
}
\email{yaodi@ict.ac.cn}
\authornote{Corresponding authors.}

\author{Chang Gong}
\affiliation{
    \institution{
    Institute of Computing Technology, Chinese Academy of Sciences\\
    University of Chinese Academy of Sciences\\
    \country{China}
    }
}
\email{gongchang21z@ict.ac.cn}

\author{Lei Zhang}
\affiliation{
    \institution{
    Strategic Data Solutions~(SDS) Group, Alibaba Inc\\
    \country{China}
    }
}
\email{zl165646@alibaba-inc.com}

\author{Sheng Chen}
\affiliation{
    \institution{
    Strategic Data Solutions~(SDS) Group, Alibaba Inc\\
    \country{China}
    }
}
\email{chensheng.cs@alibaba-inc.com}

\author{Jingping Bi}
\affiliation{
    \institution{
    Institute of Computing Technology, Chinese Academy of Sciences\\
    \country{China}
    }
}
\email{bjp@ict.ac.cn}
\authornotemark[1]

\renewcommand{\shortauthors}{Di Yao et al.}
\renewcommand{\algorithmicrequire}{ \textbf{Input:}} 
\renewcommand{\algorithmicensure}{ \textbf{Output:}} 

\newcommand{\ie}{\emph{i.e.}\xspace} 
\newcommand{\etal}{\emph{et~al.}\xspace} 
\newcommand{\etc}{\emph{etc.}\xspace} 
\newcommand{\eg}{\emph{e.g.}\xspace} 
\newcommand{\bm}{\mathbf}
\newcommand{\define}[3]{\vspace{1ex}\noindent{ \textbf{\textsc{Definition {#1}}} (#2): \emph{#3}\vspace{1ex}}
}
\newcommand\tab[1][0.5cm]{\hspace*{#1}}
\newcommand{\our}{{CausalMTA}\xspace}
\newcommand{\ourrw}{\textsc{CM-rw}\xspace}
\newcommand{\ourcasual}{\textsc{CM-causal}\xspace}
\newcommand{\ourall}{\textsc{CM-All}\xspace}
\newcommand{\red}[1]{\textcolor{red}{#1}}
\newcommand{\blue}[1]{\textcolor{blue}{#1}}
\newcommand{\ublstm}{\textsc{LSTM-ub}\xspace}

\begin{abstract}
  Multi-touch attribution (MTA), aiming to estimate the contribution of each advertisement touchpoint in conversion journeys, is essential for budget allocation and automatically advertising.
  Existing methods first train a model to predict the conversion probability of the advertisement journeys with historical data and calculate the attribution of each touchpoint by using the results counterfactual predictions. 
  An assumption of these works is the conversion prediction model is unbiased. It can give accurate predictions on any randomly assigned journey, including both the factual and counterfactual ones. Nevertheless, this assumption does not always hold as the user preferences act as the common cause for both ad generation and user conversion, involving the confounding bias and leading to an out-of-distribution (OOD) problem in the counterfactual prediction. In this paper, we define the causal MTA task and propose \our to solve this problem. It systemically eliminates the confounding bias from both static and dynamic perspectives and learn an unbiased conversion prediction model using historical data. We also provide a theoretical analysis to prove the effectiveness of \our with sufficient ad journeys. Extensive experiments on both synthetic and real data in Alibaba advertising platform show that \our can not only achieve better prediction performance than the state-of-the-art method but also generate meaningful attribution credits across different advertising channels.
  \end{abstract}



\maketitle
\section{Introduction}
\begin{figure}
    \centering
	\includegraphics[width=0.42\textwidth]{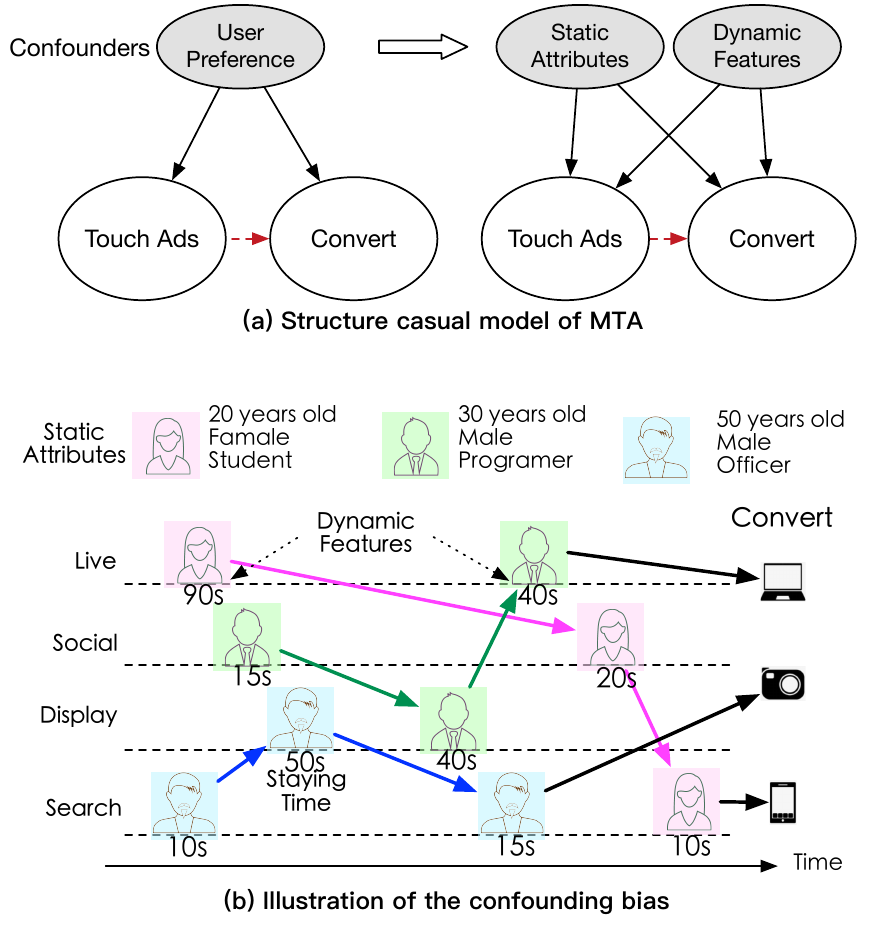}
	\vspace{-3ex}
	\caption{The motivation of \our. (a) shows the structure causal model of MTA problem. (b) illustrates the decomposition of user preference resulting in the confounding bias.}
	\vspace{-5ex}
	\label{fig:problem}
\end{figure}

Online advertising platforms have been widely deployed to help advertisers launch their advertisements (ads) across multiple marketing channels, such as social media, feed stream, and paid search. During the usage, the ad exposure sequences and conversion feedbacks of all customers are collected. Multi-touch attribution, short for MTA, aims to estimate each ad touchpoint's relative contribution in user conversion journeys. The attribution results will shed light on the budget allocation and automatically advertising. 

Artificial intelligence (AI) coupled with promising machine learning (ML) techniques well known from computer science is broadly affecting many aspects of various fields including science and technology, industry, and even our day-to-day life \cite{XU2021100179}.
Nowadays, instead of attributing the ad touchpoints by heuristic rules \cite{berman2018beyond}, data-driven methods \cite{ShaoL11,dalessandro2012causally,ji2017additional,RenFZLLZYW18,arava2018deep,DeepMTA} which estimate the attribution credits according to the historical data have become the mainstream techniques. These methods learn a conversion prediction model with all observed historical data and then generate the counterfactual ad journeys by removing or replacing some touchpoints. 
The attribution credits can be estimated using the prediction results of these counterfactual journeys based on some criteria, such as the Shapley value \cite{shapley1953value}.
One essential assumption of these methods is the conversion prediction model should be unbiased, which means the model can give fair predictions on any randomly assigned journeys, including the factual and counterfactual ones. Unfortunately, this assumption does not hold in online advertising. 

As shown in Figure \ref{fig:problem}(a), the user preference is the common cause of both `touch ads' and `convert'. Specifically, the ad exposures are recommended according to the user preferences while the user preference can also lead to the conversion. 
This common cause is a confounder in MTA, which involves spurious correlation in observed data, \ie, a connection between ads and conversion that appears to be causal but not.
As a result, the learned conversion prediction model is biased.
The discrepancy between observed training data and counterfactual data causes an out-of-distribution (OOD) problem in counterfactual prediction, which would harm the fairness of attribution. Thus, removing the influence of confounders is critical and necessary for MTA. We define the attribution of the ad journeys with an unbiased prediction model as causal MTA.

Nevertheless, it is not trivial to eliminate the confounding bias of user preferences in MTA. The reasons are two folds: (1)\textbf{ Multiple confounders.} As illustrated in Figure \ref{fig:problem} (a), the confounders in ad exposure generation consist of the static user attributes, such as genders, ages and education background, and dynamic features, \eg, previously viewed ads and favorite items. Both the static and dynamic features should be taken into account for unbiased causal MTA. Existing works either focus on the static settings \cite{austin2011introduction,johansson2016learning,johansson2018learning,zou2020counterfactual} using IPW and propensity score matching method for deconfounding, or are dedicated to the dynamic confounders \cite{Lim18a,BicaAJS20} learning an unbiased representation for prediction at each time step. All these works rely on the {strong ignorability assumption} \cite{pearl2009causality}, \ie , no hidden confounders. In their settings, the static and dynamic features are hidden confounders mutually that disable the usage. (2)\textbf{ Delay feedback.} The conversion results are observed at the end of the journey. Unlike those tasks such as \cite{CIKM17-YaoZHB17}, there is no explicit feedback available at each touchpoint. Existing sequential deconfounding methods \cite{xu2016bayesian,roy2017bayesian,Lim18a,BicaAJS20} are designed for instant feedbacks, \eg, the blood pressure, which can be observed immediately after taking the hypotensor. CAMTA \cite{CAMTA} is the most related method of our work. However, it takes the click action as the "pseudo" feedback at each touchpoint, which would involve other confounders. Above all, due to the peculiarities of advertising, there are no existed methods that can be used for unbiased causal MTA.

In this paper, we propose a novel method, namely \our, to mitigate the effect of user preferences-related confoundedness and achieve causal MTA. It learns an unbiased counterfactual prediction model which systemically eliminates the confounding bias from both static user attributes and dynamic features. One fundamental assumption of \our is that the influence of static user attributes and dynamic features are independent. This assumption is reasonable in online advertising because user attributes usually determine their item interests, and dynamic features determine how likely the users want to buy. As shown in Figure \ref{fig:problem} (b), twenty years old students tend to be attracted by fancy phones and cosmetics, whereas the middle age guys usually like high cost-performance phones and anti-bald goods. Dynamic features, such as previously visited ads and staying time, reflect the purchase intention. The main contributions can be summarized as follows: 
\begin{itemize}[leftmargin=4mm]
	\item We decompose the confounding bias of user preferences into static user attributes and dynamic features, and define the causal MTA problem. 
	\item We propose the first method \our for causal MTA, which is provable for eliminating the confounding bias of user preferences in counterfactual prediction. 
	\item Extensive experiments on a synthetic dataset, an open-source dataset and a real-world dataset of mobile phones shops from Alibaba demonstrate \our's superiority.
\end{itemize}

\vspace{-4ex}
\section{Related Work}
Existing works can be categorized into two  orthogonal groups, \ie data-driven MTA and counterfactual prediction.

\textbf{Data-driven multi-touch attribution.} Previously, marketers have applied simple rules, \eg, the last touch, to attribute the influence of touched ads \cite{berman2018beyond}, which either ignore the effects of other channels or neglect the channel difference. To overcome these drawbacks, researchers proposed data-driven methods. The data-driven MTA model was first proposed in \cite{ShaoL11}, and has been combined with survival analysis \cite{zhang2014multi} and hazard rate \cite{ji2017additional} to reflect the influence of ad exposure.
However, the data-driven methods mentioned above neglect the customers' features and cannot directly allocate personalized attribution. Besides, the temporal dependency and dynamic interaction between channels need to be modeled. Recently, many DNN-based data-driven MTA methods have been proposed to address the issues, such as channel interaction, time dependency, user characteristics.
In some studies \cite{arava2018deep,RenFZLLZYW18,du2019causally,CAMTA,DeepMTA}, RNNs are used to model longitudinal data.
DNAMTA \cite{arava2018deep} is an LSTM based deep sequential model which captures the touchpoint contextual dependency via attention mechanism and incorporates user context information and survival time-decay functions. DARNN \cite{RenFZLLZYW18} is a dual attention model that combines post-view and post-click attribution patterns for final conversion estimation. 

\textbf{Counterfactual Prediction.} Positioned as a causal estimation problem by \cite{dalessandro2012causally}, the calculation of attribution credits is actually based on counterfactual estimation \cite{zhang2014multi, du2019causally,  singal2019shapley,shender2020time}.
A limitation of the models mentioned above is the lack of exogenous variation in user exposure to advertising, which hazards the reliability of the attribution results as the training data of the counterfactual predictor is biased by confounders. Albeit extant papers \cite{barajas2016experimental, nair2017big} mitigate the issue with full or quasi-randomization, the cost and complexity of such randomization restrict the number of users and ad-types. The idea of calibrating the conversion prediction model in MTA by removing confounding bias is inspired by works in counterfactual prediction. There is a large number of methods for counterfactual prediction using observational data in the static setting, involving utilizing propensity score matching \cite{austin2011introduction}, learning unbiased representation for prediction \cite{johansson2016learning,johansson2018learning,zou2020counterfactual}, conducting propensity-aware hyperparameter tuning \cite{X17a,AlaaS18}. For estimating the effects of time-varying treatments in the area such as epidemiology where the treatments have instant feedback, many approaches \cite{xu2016bayesian,roy2017bayesian,Lim18a,BicaAJS20} addressing the longitudinal setting are proposed. Because of the gap between the longitudinal data in epidemiology and ad journeys in MTA, those methods cannot be directly used in our task.

\begin{figure*}
	\centering
	\includegraphics[width=0.9\textwidth]{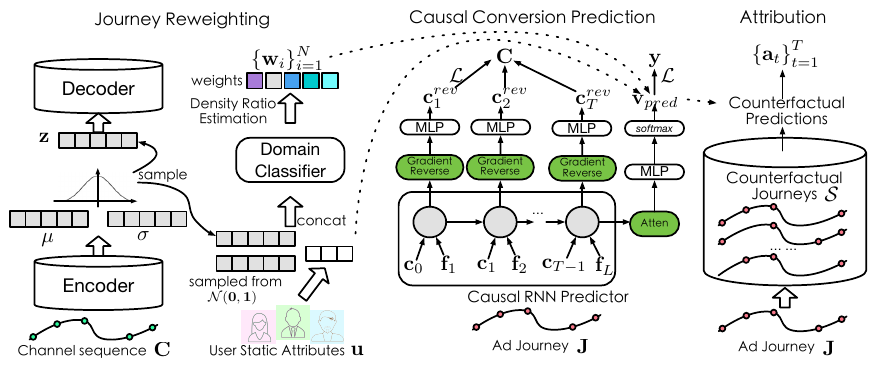}
	\vspace{-3ex}
	\caption{Architecture of \our.}
	\label{fig:figure1}
	\vspace{-3ex}
\end{figure*}
\vspace{-3ex}
\section{Preliminary}

\subsection{Problem Definition.}
We consider an ad exposure dataset ${\mathcal D}$ which consists of $N$ conversion journeys of $U$ users. Each journey can be formulated as a triplet, \ie , $( {\bm u}^{i}, {\bm J}^i, y^{i})$. 
${\bm u}^{i}$ stands for the static user attributes, which is unlikely to be changed during the user conversion journey. 
${\bm J}^i$ is a sequence of touchpoints, \ie , $\{ {\bm p}_t^{i} \}_{t=1}^{T^{i}}$. Each touchpoint ${{\bm p}^{i}_t = ({\bm c}_t^{i}, {\bm f}_t^{i})}$ contains a channel index ${\bm c}_t^{i}$ and a feature vector ${\bm z}_t^{i}$. 
Specifically, ${\bm c}_t^{i} \in \{\bm{c}_1,...,\bm{c}_k,...,\bm{c}_K\}$ indicates the exposed channels, where $K$ is the number of ad channels. The feature vector ${\bm f}_t^{i}$ includes the dynamic side information of this touchpoint, \eg , advertising form and staying time.
$y^i$ is a binary indicator that records whether the journey leads to a conversion event or not. The goal of MTA is to model the sequential pattern and assign the attribution credits to all the touchpoints ${\bm p}_t^{i}$ according to the whole information in $\mathcal{D}$. Nevertheless, historical data in $\mathcal{D}$ often exhibits confounding bias due to user preferences, which could be a fatal challenge for estimating the attribution credits. The choice of the channel at a touch-point is likely to be influenced by multiple factors like user attributes and previously visited goods. Causal multi-touch attribution aims to estimate unbiased attribution credits $\{ {\bm p}_t^{i} \}_{t=1}^{T^{i}}$ of all touchpoints.

\vspace{-3ex}
\subsection{Method Overview.}
As shown in Figure \ref{fig:figure1}, \our is a novel model-agnostic framework consisting of two key modules, \ie , journey reweighting and causal conversion prediction, which mitigate the confounding bias of user static attributes and dynamic features respectively. In journey reweighting, we employ the Variational Recurrent Auto-encoders (VRAE) to learn the generation probabilities of pure channels journeys, and conduct user demographic-based density estimation to calculate the likelihoods of the channels being randomly assigned that is used for weights computation. For causal conversion prediction, \our utilizes RNNs to model the dynamic features of journeys. A gradient reverse layer is built upon the outputs of each time step to ensure the model is unable to predict the next ad channel. 
It derives balancing representation, which removes the association between dynamic features and the ad exposure. 
The last hidden output is trained to estimate the conversion probability using the learnt weights of journey reweighting.
After that, we can obtain an unbiased prediction model. Lastly, with the constructed counterfactual journeys, the attribution credits can be estimated under Shapley value measure.

\section{Methodology}

In this section, we first specify the journey reweighting and causal conversion prediction respectively. After that, the calculation of attribution credits is detailed. In the end, we provide the theoretical analysis of \our.

\subsection{Journey Reweighting}
To mitigate the bias of static user attributes, the journey reweighting module takes pure channel sequences in ${\mathcal D}$ as the input and estimates the sample weights of the prediction model according to how likely the journey be generated randomly. It consists of two procedures, \ie , generation model for channel sequences and weights estimation of journeys.

\textbf{Generation Model for Channel Sequences.}
We utilize VRAE (Variational Recurrent Auto-encoders) \cite{FabiusAK14} to model the generation of channel sequences. 
When there is enough training data, the distribution of pure channel sequences tends to be random, regardless of user preferences.
In this setting, the learned VRAE is capable of generating unbiased predictions of observed channel sequences. 

For each ad journey $( {\bm u}, {\bm J}, y)$ in ${\mathcal D}$, we only concern with the channel information in this procedure and extract the pure channel sequence $ {\bm C} = \{ {\bm c}_t \}_{t=1}^{T}$. Taking $ {\bm C}$ as the input, \our employs the channel embedding affiliated with LSTM as the encoder and utilizes the final hidden state to generate the distribution over latent representation:
\begin{equation*}
	\begin{aligned}
	\{ {\bm h}_t \}_{t=1}^{T} &= \rm{LSTM}_{enc}({\bm C} , {\bm h}_{0}), \\
	{\bm \mu}_z &= W_{\mu}{\bm h}_{T} + {\bm b}_{\mu}, \\
	log({\sigma}_z)&=W_{\sigma}{\bm h}_{T}+{\bm b}_{\sigma},
	\end{aligned}
\end{equation*}
where $ {\bm h}_{0}$ is the initial hidden state of the encoder. Leveraging the reparametrization trick, we sample a vector ${\bm z}$ from the distribution to initialize the hidden state of the decoder:
\begin{equation*}
 	\begin{aligned}
	{\bm h'}_0 &= tanh(W^T_z{\bm z}+{\bm b}_z) ,\\
	\{{\bm h'}_t \}_{t=1}^{T}&=\rm{LSTM}_{dec}({\bm C}_{out}, {\bm h'}_0) ,\\
	{\bm c'}_{t}&=softmax(W_{o}{\bm h'}_{t}+{\bm b}_{o}) ,
	\end{aligned}
\end{equation*}
where ${\bm h'}_0$ is the initial hidden state of the decoder; ${\bm C}_{out}$ is the feed previous input which takes the output of previous step as the input; ${\bm c'}_{t}$ is the decoded channel sequence.

The loss function is composed of two parts: 1) the reconstruction loss which is defined as the cross-entropy between ${\bm c}_{t}$ and ${\bm c'}_{t}$. 2) the KL divergence between the posterior and prior distribution over the latent variable: 
\begin{equation}
	{\mathcal L}_w={\alpha}\sum_{i=1}^{N}\sum_{t=1}^{Ti}CE({\bm c}_{t},{\bm c'}_{t}) + {\beta}D_{KL}( q_{\phi}({\bm z}) || p_{\theta}({\bm z})) ,
	\label{loss:vae}
\end{equation}
where $p_{\theta}({\bm z})$ is the prior distribution usually assumed to be a standard normal distribution ${\mathcal N}(0, {\bm I})$; $q_{\phi}({\bm z}|{\bm c}^{i})$ is the posterior approximation (${\mathcal N}({\bm \mu}^{i},({\bm \sigma}^{i})^{2})$; ${\alpha}$ and ${\beta}$ are hyperparameters that control the importance each parts.

\textbf{Weights Estimation for Ad Journeys.}
To eliminate the bias of user static features, we estimate the weights of observed journeys. The journeys approximating to randomly assigned have higher  weights in conversion prediction training than those are severely affected by user preferences. Formally, the learned weights $W_T({\bm u}, {\bm C})$ should be subject to $W_T({\bm u}, {\bm C})=p({\bm C})/p({\bm C}|{\bm u})$ \cite{fong2018covariate,zou2020counterfactual}. When we learn a variational distribution $q_{\phi}({\bm z}|{\bm c})$, the variational sample weights can be computed as follows:
\begin{equation}
	\label{sample:weights}
	w^{i}=  W_T({\bm u}^{i}, {\bm c}^{i}) = \{ {\mathbb E}_{{\bm z}\sim q_{\phi}({\bm z}|{\bm c}^{i})} \ [\frac{1}{W_z({\bm u}^{i},{\bm z})}] \}^{-1} ,
\end{equation}
where $W_z({\bm u, {\bm z}})$ can be viewed as the density ratio estimation to decorrelate ${\bm u}$ and ${\bm z}$ for points in space ${\mathcal u}\times{\mathcal Z}$. 
The detailed proof can be found in the appendix.

In \our, we design a binary domain classifier to help estimate $W_ Z({\bm u},{\bm z})$. Training data of the classifier is generated cooperating with the encoder of VRAE. We label static user attributes with real latent representation $\{({\bm u}^{i},{\bm z})\}_{1\leq i \leq N}$, ${\bm z}\sim q_{\phi}({\bm z}|{\bm c}^{i})$ as positive ones, whereas samples with latent representationt sampled from standard normal distribution $\{({\bm u}^{i},{\bm z})\}_{1\leq i \leq N}$, ${\bm z}\sim p_{\theta}({\bm z})$ as negative ones. We first embed the user attributes into latent vectors and train a domain classifier to fit these samples:
\begin{equation*}
	\begin{aligned}
	\bm e_{u} &= \rm{Embedding}(\bm u)  ,\\
	\bm x &= concat(\bm e_{u}, \bm z)   ,\\
	p_{\theta_d}(L|{\bm u},{\bm z})&=sigmoid(\rm{MLP}(\bm x)) .
	\end{aligned}
\end{equation*}
Now that we have  $p(L=0)=p(L=1)$, the density ratio estimation $W_z({\bm u},{\bm z})$ can be conducted as follows:
\begin{equation}
	\label{density:ratio}
	W_z({\bm u},{\bm z}) = \frac{p({\bm u},{\bm z}|L=0)}{p({\bm u, {\bm z}}|L=1)} =\frac{p(L=0|{\bm u},{\bm z})}{p(L=1|{\bm u},{\bm z})} ,
\end{equation}
Using this formula, we can obtain the weights of all journeys, \ie , $\{{\bm w}_i \}_{i=1}^{N}$.

\subsection{Causal Conversion Prediction}
After the generation of sample weights, \our utilizes them to train a trustworthy conversion prediction model. Besides the static attributes, the biases of prediction are also caused by dynamic user features. To mitigate them, we borrow the idea from CRN \cite{BicaAJS20} and involve a gradient reverse layer to learn balancing representation.
 
Due to the delay feedback problem, we refine the structure of CRN to make it suitable for MTA. 

Formally, we first reorganize the dataset. For each journey $( {\bm u}, {\bm J}, y)$, we adopt one step offset on the channel sequence and fill the blank position with a unified placeholder, \ie ,  ${\bm C}_{+} = \{ {\bm c}_0, {\bm c}_t \}_{t=1}^{T-1}$. \our takes ${\bm C}_{+}$ along with other dynamic features ${\bm F} = \{ {\bm f}_t \}_{t=1}^{T}$ as the input and employs LSTM with the attention mechanism to obtain the trustworthy prediction: 
\begin{equation*}
	\begin{aligned}
	\bm e_{u}, \bm e_{C_+}, \bm e_{F} &= \rm{Embedding}(\bm u, {\bm C}_{+}, {\bm F}) ,\\
	\bm v_{in} &= concat(\bm e_{C_+}, \bm e_{F}) ,\\
	\{ {\bm {out}_t} \}_{t=1}^{T} &= \rm{LSTM}_{pred}(\bm v_{in}, \bm h_0) ,\\
	\end{aligned}
\end{equation*}
where $\bm e_{C_+}, \bm e_{F}, \bm v_{in}$ are the sequences of latent vectors. 
Once the output vectors are generated, we adopt them for two parallel processes. One for eliminating the bias of dynamic features, and the other for conversion prediction. 
\begin{equation*}
	\begin{aligned}
	\{ {\bm v}_t^{rev} \}_{t=1}^{T} & =\rm{MLP}(\rm{GRL}(\{ {\bm {out}}_t \}_{t=1}^{T}))  ,\\
	\{ {\bm c}_t^{rev} \}_{t=1}^{T} &= softmax(\{ {\bm v}_t^{rev} \}_{t=1}^{T})  ,\\
	\bm v_{attn} &= \rm{Attention} ({\bm {out}_T}, \{ {\bm {out}_t} \}_{t=1}^{T})   ,\\
	\bm v_{pred} &=softmax(\rm{MLP}(v_{attn}))  ,
	\end{aligned}
\end{equation*}
where $\rm GRL$ is the gradient reverse layer that ensures $\bm{out}_t$ can not predict $\bm c_t$. The loss function of causal conversion prediction consists of two parts, \ie, reverse channel prediction and conversion prediction:
\begin{equation}
	{\mathcal L}_p={\gamma}\sum_{i=1}^{N}\sum_{t=1}^{Ti}CE({\bm c}_t^{rev},{\bm c}_{t}) + {\delta }\sum_{i=1}^{N}{\bm w}_i\cdot CE({\bm v}_{pred}^i,y^i)  ,
	\label{loss:pred}
\end{equation}
where $CE$ is the cross-entropy loss; $\gamma$ and $\delta$ are hyperparameters; ${\bm w}_i $ is the learned journey weights. With the well-trained conversion prediction model, we can calculate the attribution credits of each touchpoint by constructing some counterfactual journeys. 

\subsection{Attribution Credits Calculation}
\our computes Shapley values \cite{shapley1953value} for ad credits allocation. 
Based on assessing the marginal contribution of each player in the game, the Shapley value method is a general credit distribution method, and it has been widely used in MTA \cite{singal2019shapley,DeepMTA} due to its advantage of having an axiomatic foundation and catering to fairness consideration.

Formally, let ${\bm J}^i\backslash \{{\bm p}_t^i\}$ denote the counterfactual ad journey by removing touchpoint ${\bm p}_t^i$.
$\mathcal{S}$ can be viewed as a subsequence of the counterfactual ad journey ${\bm J}^i\backslash \{{\bm p}_t^i\}$.
If we denote the result of causal conversion prediction for channel sequence ${\bm J}^i$ as $p({\bm J}^i)$,
the Shapley values for ad exposure ${\{{\bm c}_t^i\}}$ can be defined as ${SV}_t^i = \sum_{\mathcal{S}\subseteq{\bm J}^i\backslash \{{\bm p}_t^i\}} \frac{|\mathcal{S}|!(|{\bm J}^i|-|\mathcal{S}|-1)!}{|{\bm J}^i|!} [p(\mathcal{S}\cup\{{\bm p}_t^i\}) - p(\mathcal{S})]$
where $|{\bm C}^i|$, $|{\bm S}|$ are the cardinalities of these sets.
If ${SV}_t^i$ is negative, we set it zero. Then we normalize all incremental scores for each ad exposure as follows,
\begin{equation}
	\label{attr:compute}
        {\bm a}_t^i = {\sigma}({SV}_t^i)/{ \sum_{t=1}^{T^i}    {\sigma}({SV}_t^i)} ,
\end{equation}
where ${\sigma}(x)=max(0,x)$ and ${\bm a}_t^i$ are the attribution credits of the corresponding ad exposures. The pseudo-code of \our is shown in Algorithm \ref{algri}.

\begin{algorithm}[t]
	\label{algri}
	\caption{Learning procedure of \our. }
	\begin{algorithmic}[1]
	 \REQUIRE ~~\\
	 The ad exposure dataset ${\mathcal D}$, \ie, $\{({\bm u}^{i}, {\bm J}^i, y^{i})\}_{i=1}^N$ ;\\ 
	 \ENSURE ~~\\
	 Attribution credits $\{{\bm{\alpha}}_t^i\}_{t=1}^{T_i}$ for touchpoints $\{ {\bm a}_t^{i} \}_{t=1}^{T^{i}}$.\\ 
	 
	 \STATE \# Generation model for channel sequences
	 \FOR {each journey $( {\bm u}^{i}, {\bm J}^i, y^{i})$ in ${\mathcal D}$}
	 \STATE Evaluate ${\mathcal L}_w$ according to Eq.(\ref{loss:vae}) and update the parameters of VRAE model.
	 \ENDFOR
	 
	 \STATE \# Weights estimation for ad journeys
	 \FOR {each ad journey $( {\bm u}^{i}, {\bm J}^i, y^{i})$ in ${\mathcal D}$}
	 \STATE Generate latent representation for positive and negative samples respectively. 
	 \STATE Optimize the parameters of domain classifier.
	 \ENDFOR
	 \STATE Conduct density ratio estimation and calculate sample weights ${\bm w}^{i}$ according to Eq.(\ref{sample:weights}) and Eq.(\ref{density:ratio}). 
	 
	 \STATE \# Causal conversion prediction
	 \FOR {each ad journey $( {\bm u}^{i}, {\bm J}^i, y^{i})$ in ${\mathcal D}$}
	 \STATE Evaluate ${\mathcal L}_p$ according to Eq.(\ref{loss:pred}) and update the parameters of conversion prediction model.
	 \ENDFOR
	 
	 \STATE \# Calculation of Attribution Credits
	 \FOR {each ad journey $( {\bm u}^{i}, {\bm J}^i, y^{i})$ in ${\mathcal D}$}
	 \FOR {each touchpoint ${\bm c}_t^i$ in channel sequence ${\bm C}^i$}
	 \STATE Compute $SV_t^i$ according to its definition.
	 \STATE Calculate attribution credits $\bm{a}_t^i$ according to Eq.(\ref{attr:compute}).
	 \ENDFOR
	 \ENDFOR
	 \RETURN $\{\{{\bm{a}}_t^i\}_{t=1}^{T_i}\}_{i=1}^N$
	\end{algorithmic}
	\label{algri}
	\vspace{-1ex}
\end{algorithm}
   
\subsection{Theoretical Analysis of \our}
Under the assumption of independence, we can decompose the overall confounding bias $\mathcal B$ into the bias introduced by user static attributes ${\mathcal B}_\bm{u}$ and the bias introduced by dynamic user features ${\mathcal B}_\bm{F}$, \ie, ${\mathcal B} = {\mathcal B}_\bm{u} + {\mathcal B}_\bm{F}$. \our aims to obtain an unbiased prediction model and achieve ${\mathcal B} = 0$.

We prove that the confounding bias from static user attributes ${\mathcal B}_\bm{u}$ can be mitigated by estimating sample weights ${\bm w}^i$ for ad journeys.
Formally, let ${\mathcal E}_{cf}$ denote the counterfactual prediction error, which is the target to be minimized. Unfortunately, ${\mathcal E}_{cf}$ can not be directly measured on the observational dataset. 
We can derive the upper bound of ${\mathcal E}_{cf}$, which is given by
\begin{equation*}
    {\mathcal B}_\bm{u} = {\mathcal E}_{cf} - {\mathcal E}_{f}^w \leq IPM_{G}\left (W_T({\bm u},{\bm C})p({\bm u},{\bm C}), \ p({\bm u})p({\bm C}) \right )  ,
\end{equation*}
where ${\mathcal E}_{f}^w$ is the prediction error on the re-weighted data and $IPM$ denotes Integral Probability Metric. 
When $W_T({\bm u},{\bm C}) = \frac{p({\bm C})}{p({\bm C}|{\bm u})}$, the equation ${\mathcal E}_{cf}={\mathcal E}_{f}^w$ can be proved.
More details of the proof are available in the appendix.

In dynamic settings, ${\mathcal B}_\bm{F}$ equals zero if we can prove that the learned representation removes the association between dynamic features and the ad exposure. 
We build the representation ${\bm v}_t^{rev}$ invariant across different ad channels to eliminate biases caused by dynamic user features. We achieve this by minimizing the formula ${\mathcal L}_{rev} = \sum_{i=1}^{N}\sum_{t=1}^{Ti}CE({\bm c}_t^{rev},{\bm c}_{t})$ in  Eq.(\ref{loss:pred}). 
We can prove that 
\begin{equation*}
{\mathcal L}_{rev}=K\cdot JSD(p({\bm v}_t^{rev}|{\bm c}_1),...,p({\bm v}_t^{rev}|{\bm c}_K))-K\ logK   ,
\end{equation*}
where $KlogK$ is a constant, and $JSD(\cdot,...,\cdot)$ denotes the multi-distribution Jensen-Shannon Divergence \cite{Li18Proof}, which is non-negative and 0 if and only if all distributions are equal. To minimize ${\mathcal L}_{rev}$, we derive $p({\bm v}_t^{rev}|{\bm c}_1)=...=p({\bm v}_t^{rev}|{\bm c}_K)$, where ${\bm v}_t^{rev}$ is the learned representation invariant across different ad channels. For details, see the appendix.

\section{Experiment}
In this section, we evaluate the performance of \our and answer the following questions:
\begin{itemize}[leftmargin=4mm]
    \item \textbf{Q1:} What is the performance of \our in terms of eliminating confounding bias?
    \item \textbf{Q2:} Does \our outperform the state-of-the-art MTA methods in conversion prediction?
    \item \textbf{Q3:} How does \our perform on real-world ad impression datasets?
    \item \textbf{Q4:} What are the capabilities of the journey re-weighting module and the causal conversion prediction module?
\end{itemize}

\begin{table*}[t]
    \centering
    \caption{Results of conversion prediction on synthetic dataset. {\textsc{MTA-ub}\xspace} is the upper-bound performance trained on unbiased data.}
    \begin{tabular}{c|cc|cc|cc}
    \toprule
    Method & AUC & Log-loss& AUC & Log-loss& AUC & Log-loss \\
    \midrule
    &\multicolumn{2}{c|}{dynamic-only} &\multicolumn{2}{c|}{static-only} &\multicolumn{2}{c}{hybrid}\\
    
    LR & 0.6256$\pm$0.02 & 0.8414$\pm$0.001 & 0.6181$\pm$0.01 & 0.8781$\pm$0.002 & 0.5883$\pm$0.03 &  1.1226$\pm$0.002 \\
    SP & 0.5731$\pm$0.00 & 1.0712$\pm$0.00 & 0.5514$\pm$0.00 & 1.3361$\pm$0.00 & 0.5210$\pm$0.00 &  1.8853$\pm$0.00 \\
    AH & 0.6328$\pm$0.01 & 0.7942$\pm$0.001 & 0.6231$\pm$0.02 & 0.8515$\pm$0.002 & 0.5832$\pm$0.01 &  1.1136$\pm$0.002 \\
    DNAMTA & 0.6497$\pm$0.02 & 0.6795$\pm$0.001 & 0.6465$\pm$0.02 & 0.6624$\pm$0.002 & 0.6147$\pm$0.03 &  0.6497$\pm$0.002 \\
    DeepMTA & 0.6519$\pm$0.02 & 0.6778$\pm$0.001 & 0.6427$\pm$0.02 & 0.6678$\pm$0.002 & 0.6073$\pm$0.03 &  0.6519$\pm$0.002 \\
    CAMTA & 0.6926$\pm$0.02 & 0.6583$\pm$0.001 & 0.6531$\pm$0.01 & 0.6927$\pm$0.002 & 0.6485$\pm$0.03 &  0.6872$\pm$0.002 \\
    \our & 0.7034$\pm$0.01 & 0.6472$\pm$0.002 & 0.6882$\pm$0.02 & 0.6521$\pm$0.003 & 0.6814$\pm$0.01 & 0.6424$\pm$0.003\\
    {\textsc{MTA-ub}\xspace} & 0.7268$\pm$0.01 & 0.6454$\pm$0.002 & 0.7205$\pm$0.01 & 0.6353$\pm$0.002 & 0.7116$\pm$0.01 &  0.6285$\pm$0.002\\
    \midrule
    \end{tabular}
    \label{tab:synthetic}
\end{table*}

\subsection{Experimental Settings}
This section provides an overview of the data, experimental protocol, evaluation metrics, compared baselines, and hyperparameter settings. More details of this part can be found in the Appendix \ref{appen:exp}. All the code and data are available in the supplementary file, and will be released after acceptance.

\subsubsection{Data Descriptions.}
On three datasets, the performance of \our is evaluated. The first dataset is a synthetic dataset created to test \our's ability to solve the issue of confounding bias. 
The second, \textbf{Criteo}$\footnote{http://ailab.criteo.com/criteo-attribution-modeling-bidding-dataset/}$, is a publicly available dataset on ad bidding that is commonly utilized in MTA \cite{Meynet_Criteo,RenFZLLZYW18,CAMTA}.
We use the same experimental setup as CAMTA\cite{CAMTA} to preprocess it.
More details of the preprocessing are specified in the Appendix \ref{appen:process}.  
The third dataset is a real ad impression dataset from \textbf{Alibaba}, which includes $30$ days of ad impression data from mobile phone shops.
These touchpoints are categorized into $40$ channels, including interact, feed, display, search, live show, \etc

\begin{table}[t]
    \centering
    \label{tab:dataset}
    \caption {The overview of the Criteo dataset} 
    \vspace{-2ex}
 \begin{tabular}{cccc}
  \toprule  
  {Statistics} & \textbf{Raw} & \textbf{Processed} \\ 
  \midrule  
  No. of users & 6,142,256 & 157,331 \\   
  No. of campaigns & 675 & 10 \\
  No. of journeys & 6,514,319 & 196,560 \\    
  No. of convert journeys & 435,810 & 19,890\\ 
  No. of touchpoints & 16,468,027 & 787,483\\  
  \bottomrule  
 \end{tabular}
 \vspace{-4ex}  
\end{table}

\subsubsection{Experiment Protocol.}
To evaluate \our's performance, we conduct experiments on three datasets.
For synthetic dataset, we simulate various confounder settings to obtain the biased and unbiased data, and we quantify the ability of \our to eliminate confounding bias quantitatively with conversion prediction (Section \ref{sec:exp_synthetic}). For Criteo dataset, 
we compare \our's performance to that of the state-of-the-art methods under two tasks, \ie, conversion prediction and data reply (Section \ref{sec:exp_criteo}).
We also report the experimental results of \our on Alibaba advertising platforms, which contains the attribution value analysis and profit comparison (Section \ref{sec:exp_ali}). Moreover, we provide the ablation studies for verifying the effectiveness of the proposed journey re-weighting and causal conversion prediction modules (Section \ref{sec:exp_abl}).

\subsubsection{Evaluation Metrics.} For conversion prediction, we evaluate the performance in terms of \textbf{log-loss}, \textbf{AUC}. For fairness, the log-loss only contains the conversion prediction part of Equation \ref{loss:pred}. It can be reckoned as a standard measurement to estimate the classification performance. AUC can be a metric reflecting the pairwise ranking performance of the estimation between converted and non-converted ad impression sequences. 

For data repaly experiments, we follow the work in \cite{RenFZLLZYW18} and utilize the \textbf{return on investment} (ROI) as the metric to compute the overall budget allocation. Then, the historical data are selected to fit the budget. We compared the performance of different budget allocation in back evaluation with two metrics, \ie, Cost per Action (\textbf{CPA}) and Conversion Rate (\textbf{CVR}). CPA is the total monetary cost normalized by the number of conversions, which measures the efficiency of advertising campaign. And CVR is the number of converted sequences averaged by number of testing sequences, which reflects the ratio of gain for the ad exposure.

\subsubsection{Compared Methods.} In our experiments, \our is compared with $8$ baseline methods which can be divided into three categories, \ie, statistical learning-based methods, deep learning-based methods, and causal learning-based methods. The statistical learning-based methods consist of three methods, \ie, Logistic Regression \cite{ShaoL11} (\textbf{LR}), Simple Probabilistic \cite{dalessandro2012causally} (\textbf{SP}), and Additive Hazard \cite{zhang2014multi} (\textbf{AH}). Deep learning-based methods contain three methods, \ie, \textbf{DNAMTA} \cite{arava2018deep}, \textbf{DARNN} \cite{RenFZLLZYW18}, and \textbf{DeepMTA} \cite{DeepMTA}. The causal learning-based methods also have two works, \ie,  \textbf{JDMTA} \cite{du2019causally} and \textbf{CAMTA} \cite{CAMTA}. Besides, we also compare \our with two ablation methods, \ie, \ourrw and \ourcasual. Detailed descriptions of these methods are available in the Appendix \ref{appen:baseline}.

\subsubsection{Parameter Settings.}  For LSTMs in \our, we stack three $3$ layer LSTMs as the encoder, decoder, and the predictor respectively. MLP models in \our are composed of $4$ fully connected layers with $ELU$ as the activate function. \our has $4$ hyperparameters \ie, $\alpha, \beta, \gamma$ and $\delta$, we empirically set $\alpha = \beta = 0.5$, and $\gamma = \delta = 0.5$. All the experiments are conducted on a high-end server with $2\times $ NVIDIA GTX3090 GPUs. All the compared baselines are trained in $30$ epochs and the best model is chosen to report.

\subsection{Experiments on Synthetic Data} \label{sec:exp_synthetic}
To answer \textbf{Q1}, we conduct the synthetic experiment. Next, we introduce the generation process of synthetic data and report the experimental results respectively. 

\subsubsection{Data Generation.} The data generation procedure is composed of the ad exposure policy and the user conversion module. For the ad exposure policy, we first generate the sequence of exposure events with a Poisson process. Then, a stochastic function is designed to assign the ad types for the events. The parameters of both Poisson process and the stochastic function are related to the user preference. As for user conversions, we follow the work in \cite{tp_MTA_shender20} which set the conversion probabilities of all ad types as a function of user demographics. The conversion probability of a specified ad journey can be calculated by aggregating the probabilities of all related ads. The detailed data generation procedure can be found in the Appendix \ref{appen:generation}.

Following the data generation process, we construct three sub-datasets, \ie, dynamic-only, static-only and hybrid, which reflect different kinds of confounding bias. For dynamic-only, the user preference only influence the generation of exposure sequences and the ad types are randomly assigned. For static-only, the user preference only influence the ad types and the sequence is generated by a fixed Poisson distribution. For hybrid, the user preference influence both. Moreover, we also construct unbiased training datasets to train prediction models and treat them as the performance upper-bound.

\subsubsection{Performance comparison.}
As shown in Table \ref{tab:synthetic}, \our significantly outperforms the compared method in three confounder settings. The performances are very close to the performance upper-bound, which indicates \our is able to alleviate the confounding bias in MTA. By considering the casual relationship, CAMTA achieves the best performance among all the competitors, but is also inferior to \our. This phenomenon verifies the importance of modeling the causal mechanism for MTA.

Comparing the results of different settings, we observe \our achieves the best performance in hybrid confounders. It proves the effectiveness of the proposed techniques. 
The performance of \our in dynamic-only and static-only settings does not degrade significantly, indicating that it's applicable to a wide range.

\begin{table}[t]
    \caption{Results of conversion prediction on Criteo dataset. SL, DL, CL in the left column indicate the statistical learning-based methods, deep learning-based methods and causal learning-based methods, respectively.}
    \label{tab:performance_comp}
    \scalebox{0.95}{
    \begin{tabular}{c|c|cc}
    \toprule
     & Method & AUC &Log-loss \\
    \midrule
    \multirow{3}{*}{SL}& LR &0.8370$\pm$0.03  &0.191$\pm$0.01 \\
    &SP & 0.7616$\pm$0& 0.317$\pm$0\\
    &AH &0.7264$\pm$0.03 &0.286$\pm$0.01\\
    \midrule
    \multirow{2}{*}{DL} &DNAMTA &0.9127$\pm$0.02 &0.1360$\pm$0.013 \\
    &DARNN &0.8726$\pm$0.02 &0.165$\pm$0.006\\
    &DeepMTA &0.9104$\pm$0.03 &0.112$\pm$0.012\\
    \midrule
    \multirow{2}{*}{CL} &JDMTA &0.9127$\pm$0.01&0.0838$\pm$0.007\\
    &CAMTA &0.9347$\pm$0.02 &0.0715$\pm$0.007 \\
    \midrule
    \multirow{3}{*}{Ours} &\our &\textbf{0.9659$\pm$0.01} &\textbf{0.0517$\pm$0.003} \\
    &\ourrw &0.9539$\pm$0.01&0.0560$\pm$0.003\\
    &\ourcasual &0.9517$\pm$0.01&0.0534$\pm$0.002\\
    \bottomrule
    \end{tabular}}
    \vspace{-4ex}
\end{table}

\subsection{Experiments on Criteo Dataset} \label{sec:exp_criteo}
To answer \textbf{Q2}, we employ the most widely used public dataset Criteo to evaluate the performance of \our. In this section, we first briefly describe the tasks and then specify the results. 

\subsubsection{Task Description}
Tow tasks, \ie conversion prediction and data replay, are conducted to test the performance of the unbiased prediction model and the attribution weights, respectively. For conversion prediction, we directly train \our and baselines under the same setting, and compare the prediction accuracy. For data replay, we follow the experiments in \cite{RenFZLLZYW18} and utilize the attribution credits to compute the \textbf{return on investment} (ROI) and budget allocation on different ads. 
Based on the result, we conduct back evaluation for budget allocation, and measure the CPA and CVR of them. 
\vspace{-3ex}

\subsubsection{Performance of Conversion Prediction}
The detailed evaluation results of conversion prediction on different baselines are presented in Table \ref{tab:performance_comp}. 
As shown, \our continuously outperforms all the compared baselines, which proves the validity of eliminating the confounding bias on static user attributes and dynamic features. CAMTA is the strongest baseline but also inferior to \our. It utilizes click labels as the auxiliary information, which probably involves additional confounding bias. Moreover, CAMTA does not consider the difference between static and dynamic confounders, which would also harm the performance. One interesting phenomenon is \our has a more stable confidence interval compared to other baselines. It indicates that the parameters in \our tend to converge to similar values with different initialization. To a certain extent, \our is more robust than other baselines. 

Comparing the performance of different categories of methods, we can observe that SL methods are inferior to the other two categories. SL methods either use statistical laws or employ logistic regression to predict the conversion probability, which can not well model the conversion process. DL methods perform better than the SL methods but are also inferior to the CL methods. It proves that the deep learning techniques are more suitable for conversion prediction due to their large parameter space and high ability to model complex tasks. However, these methods have poor performance compared to the CL methods. It is because deep learning methods directly use the observed data to train the prediction models, which are incapable of handling confounding bias and would suffer from the out-of-distribution problem. CL methods outperform other baselines with a large margin, which demonstrates the prediction performance highly increased by eliminating the confounding bias.

\subsubsection{Performance of Data Replay}
In this experiment, we evaluate \our and the compared baselines under four proportions of total budgets, \ie, $1/2$, $1/4$, $1/8$, $1/16$. Following the setting of \cite{CAMTA}, we use value $V(y^i)=1$ for computing $\mathrm{ROI}_{{c}_k}$ and scale the cost by $1000$ to highlight the difference of CPA in the comparison. The detailed results are shown in Table \ref{tab:data_repaly}. We can observe that: (1) In most cases, \our achieves the best performance, which verifies the effectiveness of the proposed methods. (2) The conversion number of DeepMTA is better than our method in $1/8$ and $1/16$ budget, but inferior to \our in $1/2$ and $1/4$ budget. This phenomenon shows that \our captures the intrinsic characters of attribution and performs better in large budget. (3) Among all the baselines, CAMTA is the strongest competitor. It indicates the performance can be improved by involving deconfounding mechanism. (4) Deep learning methods significantly outperform traditional methods. Benefiting from the high complexity, deep learning techniques are more suitable for modelling the user conversion.

\begin{table*}[t]
    \centering
    \caption{The results of data reply experiment on Criteo dataset. CPA: the lower the better; CVR: the higher the better.}
    \begin{tabular}{c|cccc|cccc|cccc}
    \toprule
    & \multicolumn{4}{c}{CPA(Cost pre action)} & \multicolumn{4}{c}{Conversion Number} & \multicolumn{4}{c}{CVR(Conversion rate)} \\
    Method & $\frac{1}{2}$ & $\frac{1}{4}$ & $\frac{1}{8}$ & $\frac{1}{16}$ & $\frac{1}{2}$ & $\frac{1}{4}$ & $\frac{1}{8}$ & $\frac{1}{16}$ & $\frac{1}{2}$ & $\frac{1}{4}$ & $\frac{1}{8}$ & $\frac{1}{16}$\\
    \midrule
    LR & 58.41 & 56.43 & 55.26 & 55.25 & 913 & 627 & 342 & 193 & 0.0835 & 0.0879 & 0.0946 & 0.0823 \\
    SP & 49.60 & 46.15 & 48.64 & 45.39 & 842 & 548 & 375 & 207 & 0.0789 & 0.0772 & 0.0914 & 0.0910 \\
    AH & 51.66 & 47.30 & 47.67 & 52.65 & 843 & 594 & 382 & 135 & 0.0804 & 0.0826 & 0.0895 & 0.0775 \\
    DNAMTA & 38.79 & 35.12 & 31.47 & 32.35 & 1181 & 778 & 459 & 264 & 0.1039 & 0.1068 & 0.1131 & 0.1148 \\
    DARNN & 32.62 & 30.46 & 28.09 & 28.72 & 1286 & 829 & 480 & 244 & 0.1218 & 0.1237 & 0.1241 & 0.1223 \\
    DeepMTA & 36.25 & 30.60 & 26.08 & 25.97 & 1372 & 880 & \textbf{549} & \textbf{289} & 0.1194 & 0.1202 & 0.1236 & 0.1249 \\
    CAMTA & 32.61 & 29.73 & \textbf{26.05} & 26.25 & 1270 & 864 & 538 & 211 & 0.1127 & 0.1160 & 0.1191 & 0.1166 \\
    \our & \textbf{30.34} & \textbf{29.52} & 26.45 & \textbf{25.47} & \textbf{1441} & \textbf{976} & 548 & 255 & \textbf{0.1247} & \textbf{0.1265} & \textbf{0.1305} & \textbf{0.1283} \\ 
    \bottomrule
    \end{tabular}
    \label{tab:data_repaly}
\end{table*}

\subsection{Empirical Applications in Alibaba}\label{sec:exp_ali}
To answer \textbf{Q3}, we evaluate the performance of \our on Alibaba advertising platform. 
Channel attributions of each shop are more meaningful to guide the budget allocation. We train the attribution models in the first $15$ days and use the last data for testing. We choose all of its converted journeys in the test set for one specific shop and compute the mean credits of $40$ channels. After that, we employ two experiments, \ie, attribution improvement and offline data replay, to evaluate the performance of \our.

\begin{figure}
    \centering
	\includegraphics[width=0.45\textwidth]{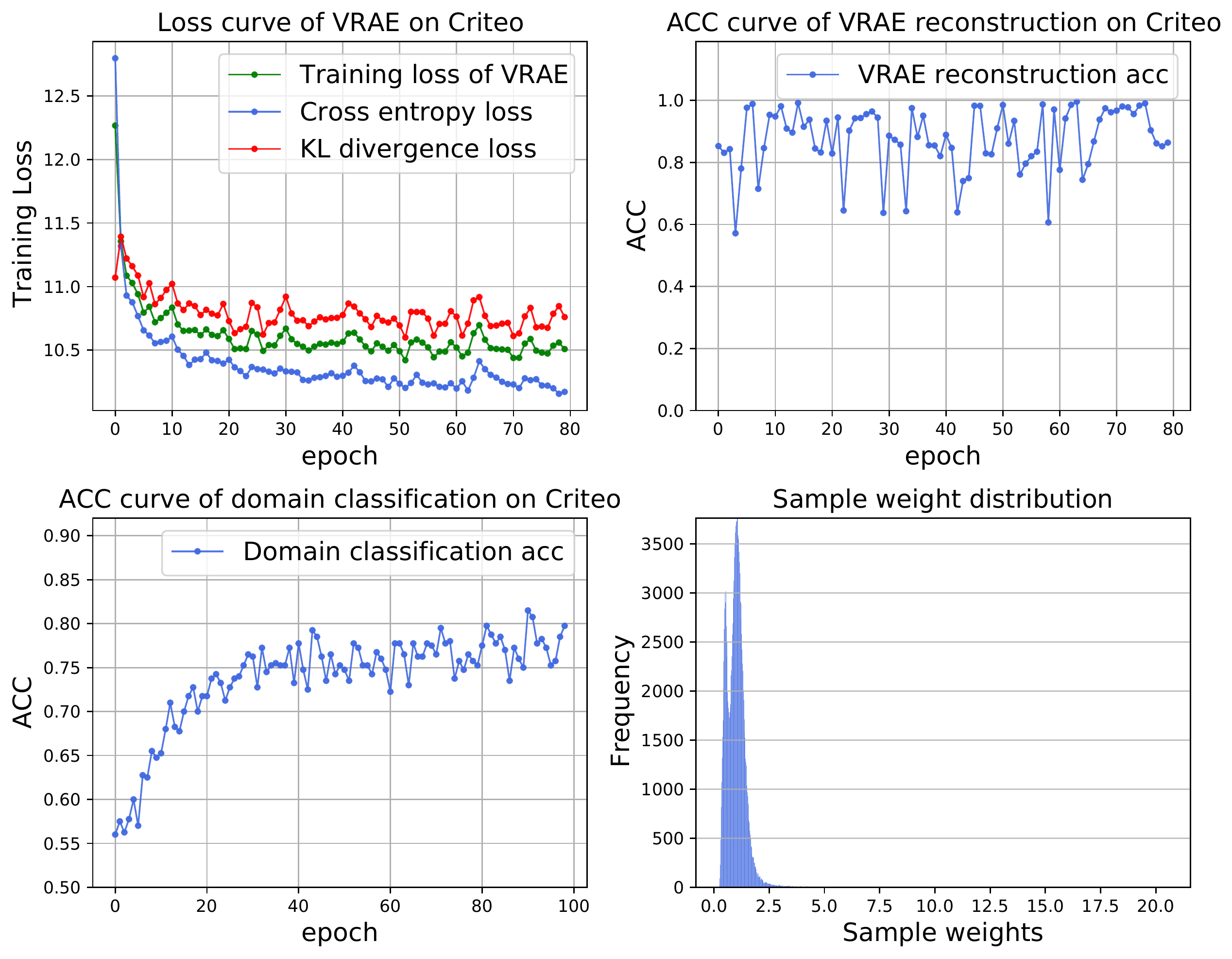}
	\caption{Learning curves and sample weight distribution on Criteo dataset.}
	\label{fig:jwablation}
    \vspace{-4ex} 
\end{figure}

\subsubsection{Attribution Improvement.} 
We compute the attribution credits of two representative cellphone shops 
utilizing \our and compare them with the credits calculated by an LSTM-based conversion prediction model. As shown in Figure \ref{fig:perfablation}, the credits on both shops decreased, indicating that the estimated contribution of search ads is reduced after eliminating the confounding bias of user preferences. This is because user tends to search the goods before paying. The attributions of the search channel are usually overestimated, and \our can mitigate this kind of bias.

\begin{figure}
    \centering
	\includegraphics[width=0.45\textwidth]{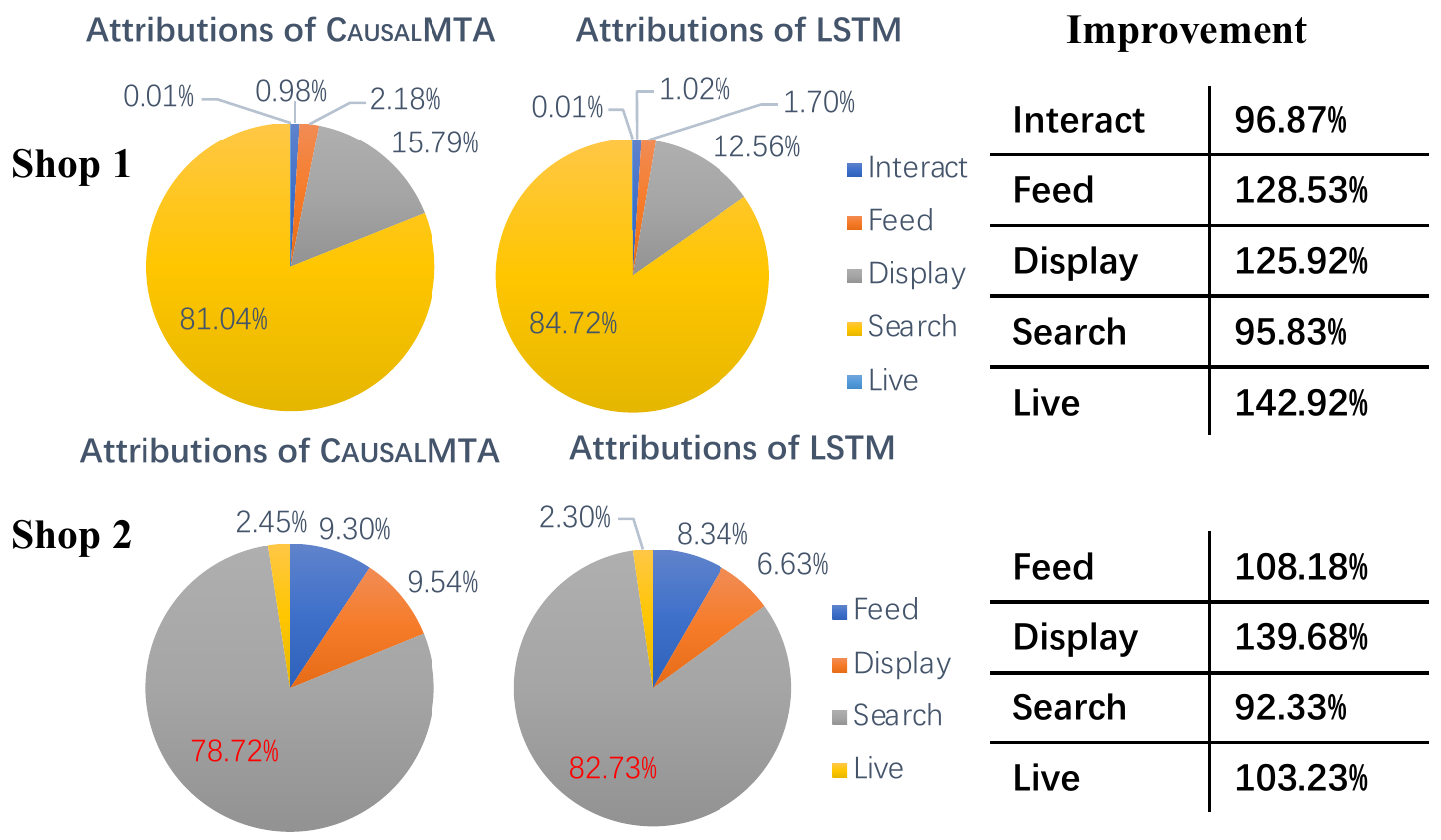}
	\caption{The comparison of attribution changes for two cellphone shops.}
	\label{fig:perfablation}
    \vspace{-4ex}
\end{figure}

\begin{table}[t]\centering
    \caption{Profit comparison of data replay.}
    \label{tab:datarepaly}
    \scalebox{0.9}{
    \begin{tabular}{c|cccc}
    \toprule
    Method & 1/2 & 1/4 &1/8 &1/16\\
    \midrule
    &\multicolumn{4}{c}{Shop 1}\\
    LSTM&69.8 &63.7 &56.2 &58.3\\
    \our&72.3 &70.2 &57.1 &59.2\\
    \midrule
    Improvement&+3.58\% &+10.20\% &+1.60\% &+1.54\%\\
    \midrule
    &\multicolumn{4}{c}{Shop 2}\\
    LSTM&20.9 &18.3 &17.5 &15.2\\
    \our&21.1 &19.1 &17.8 &15.8\\
    \midrule
    Improvement&+0.96\% &+4.37\% &+1.71\% &+3.95\%\\
    \bottomrule
    \end{tabular}}
	\vspace{-3ex}
\end{table}

\subsubsection{Performance of Data Replay.} 
In this experiment, we employ the attribution credits to guide the budget allocation. Based on the attribution credits, we first compute the return-on-investment (ROI) of each channel and utilize the normalized weights of ROI as budget proportion \cite{RenFZLLZYW18}. Assuming that the total cost of the test set is $cost_t$, we set the evaluation budgets as $1/2, 1/4, 1/8, 1/16$ of $cost_t$ and replay the historical data to select journeys satisfying evaluation budgets. Table \ref{tab:datarepaly} shows the comparison results of the profit in each evaluation budget. We can observe that the profit \our consistently outperforms the LSTM-based predictor on all evaluation budgets, which indicates that the attribution credits of \our reflect the causal relationships in advertising. It can be used to guide budget allocation and achieve better profit.

\begin{figure}
    \centering
	\includegraphics[width=0.45\textwidth]{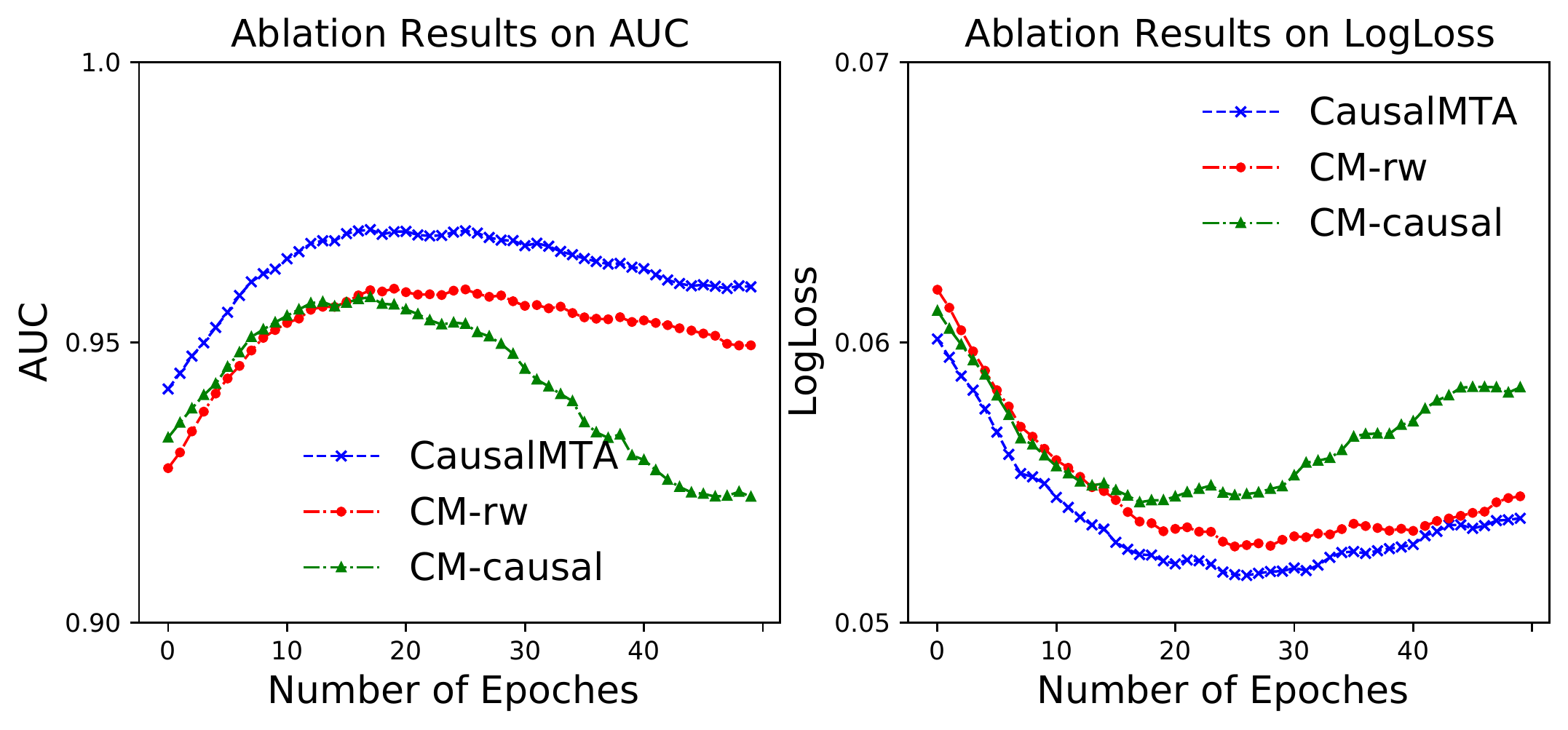}
    \vspace{-2ex}
	\caption{Performance comparison of ablation methods.}
	\label{fig:perfablation}
    \vspace{-6ex}
\end{figure}
\subsection{Ablation Studies} \label{sec:exp_abl}
To explore the effectiveness of the proposed methods, we compare \our with two ablation methods on Cirteo dataset, \ie, \ourrw and \ourcasual, which remove the reweighting procedure and the gradient reverse layer respectively. 
We first show the intermediate results of journey reweighting. 
The reconstruction accuracy of VRAE and the classification accuracy of the domain classifier directly influence the performance of \our. As shown in Figure \ref{fig:jwablation}, the reconstruction accuracy is approximate to $98\%$, and the classification accuracy is approximate to $82\%$, which indicates that VRAE and domain classifier are well trained, and results of journey reweighting are significant. 
We also witness the reconstruction accuracy fluctuates at a high level as KL divergence dominates the training loss when the cross-entropy loss is small enough.

We summarize the metrics of ablations in Table \ref{tab:performance_comp} and illustrate the training procedure in Figure \ref{fig:perfablation}. As shown, the AUCs of \ourrw and \ourcasual are inferior to \our, which proves that both the journey reweighting and causal conversion prediction help improve the performance. By removing the journey reweighting model, the performance of \our decreases from $0.9659$ to $0.9539$. By removing the gradient reverse layer, the performance of \our decreases from $0.9659$ to $0.9617$. We can observe the improvement of gradient reverse layer is more significant than journey reweighting, which indicates the confounding bias of dynamic feature are more obvious than the static user attributes. This result is consistent with our cognition. Moreover, as illustrated in Figure \ref{fig:perfablation}, the convergence speed of \our is faster than \ourrw and \ourcasual, which shows the superiority of \our in eliminating the confounding bias of user preferences.

\section {Conclusion}
In this paper, we define the problem of causal MTA, which eliminates the confounding bias introduced by user preferences and assigns the attribution credits fairly over all touchpoints. We propose \our, which decomposes the confounding bias of user preferences into two independent parts, \ie, the static user attributes and the dynamic features. \our employs journey reweighting and causal conversion prediction to solve these two kinds of confounding bias respectively. We prove that \our is capable of generating unbiased conversion predictions of ad journey. Extensive experiments on the public dataset and real commercial data from Alibaba show that \our outperforms all the compared baselines and works well in the real-world application.

\begin{acks}
    This work has been supported by the National Natural Science Foundation of China under Grant No.: 62002343, 62077044, 61702470.
\end{acks}

\bibliographystyle{ACM-Reference-Format}
\bibliography{references}
\clearpage
\appendix
\appendix

\section{Appendix}
\subsection{Theoretical Analysis of \our}\label{appen:theory}

\subsubsection{Proof of weights calculation.}
In order to create a pseudo-population which debiases by means of sample re-weighting, the weights should cater to the equation $W_T({\bm U}, {\bm C})=p({\bm C})/p({\bm C}|{\bm U})$ \cite{fong2018covariate,zou2020counterfactual}. When we learn a variational distribution $q_{\phi}({\bm z}|{\bm c})$ of the original channel assignment of touch-points, the variational sample weights can be computed as follows:
\begin{align*}
w^{i}&=  W_T({\bm u}^{i}, {\bm c}^{i})  = \frac{p({\bm c}^{i})}{p({\bm c}^{i}|{\bm u}^{i})} \\
&=\frac{p({\bm c}^{i})}{   \int_{\bm z} p({\bm c}^{i}|{\bm z})  \ p({\bm z} | {\bm u}^{i})  \ d{\bm z}    } = \frac{1}{ \int_{\bm z} \frac{p({\bm c}^{i}|{\bm z})}{p({\bm c}^{i})}  \   p({\bm z}|{\bm u}^{i})   \     d{\bm z}}\\
&= \frac{1}{ \int_{\bm z} \frac{p({\bm z}|{\bm c}^{i})}{p({\bm z})}  \   p({\bm z}|{\bm u}^{i})   \     d{\bm z}} = \frac{1}{ \int_{\bm z} \frac{p({\bm z}|{\bm u}^{i})}{p({\bm z})}  \   p({\bm z}|{\bm c}^{i})   \     d{\bm z}}\\
&= \frac{1}{ \int_{\bm z} \frac{p({\bm z},{\bm u}^{i})}{p({\bm z})\ p({\bm u}^{i})}  \   p({\bm z}|{\bm c}^{i})   \     d{\bm z}} = \frac{1}{ \int_{\bm z}  \frac{1}{W_Z({\bm u}^{i},{\bm z})}    \   p({\bm z}|{\bm c}^{i})   \     d{\bm z}}\\
&= \frac{1}{ {\mathbb E}_{{\bm z}\sim q_{\phi}({\bm z}|{\bm c}^{i})} \ [\frac{1}{W_Z({\bm u}^{i},{\bm z})}]  }  ,
\end{align*}
where $W_Z({\bm U, {\bm Z}})$ can be viewed as the density ratio estimation to decorrelate ${\bm U}$ and ${\bm Z}$ for points in space ${\mathcal U}\times{\mathcal Z}$. 

\subsubsection{Proof of the journal reweighting module.}

In the task of multi-touch attribution, provided with observational data, we hope to learn a hypothesis $f_{\theta_p}:{\mathcal U}\times{\mathcal C}\mapsto{\mathbb R}$ with model parameters $\theta_p$, which predicts the conversion rate based on the confounders and touch-points. In this setting, the concept of counterfactual is to guarantee the learned hypothesis to predict accurate outcome when the assignment of touch-point (e.g., the channel preference) is random. For the individual ${\bm U}$, when ${\mathcal L}()$ denotes the error function and $y()$ denotes the ground-truth outcome, the prediction error can be formed as:
\begin{equation*}
    {\mathcal E}({\bm U})={\mathbb E}_{{\bm C}\sim p({\bm C})} 
    \left[
    {{\mathcal L}\left(
    f_{\theta_p}({\bm U},{\bm C}), y({\bm U},{\bm C})
    \right)}
    \right]  .
\end{equation*}
The target of the counterfactual prediction error to be minimized is ${\mathcal E}_{cf}={\mathbb E}_{{\bm U}\sim p({\bm U})}[{\mathcal E}({\bm U})]$. But in the observational dataset, the touch-points are assigned based on confounders, i.e., ${\bm C}\sim p({\bm C}|{\bm U})$. Instead of directly using supervised learning, optimizing the prediction error on the re-weighted data
\begin{equation*}
{\mathcal E}_{f}^w={\mathbb E}_{{\bm U},{\bm C}\sim p({\bm U},{\bm C})  } 
\left[
{{\mathcal L}
\left(f_{\theta_p}({\bm U},{\bm C}), y({\bm U},{\bm C})\right)} W_T({\bm U},{\bm C}) 
\right]  ,
\end{equation*}
can lead to a more accuracy counterfactual prediction.

Assuming a family $G$ of functions $g:{\mathcal U}\times{\mathcal C}\mapsto{\mathbb R}$, and we have ${\mathcal L}(f({\bm U},{\bm C}), y({\bm U}, {\bm C})) = l({\bm U},{\bm C}) \in G$ . We can therefore bridge the gap between the counterfactual loss and the re-weighted loss under observational data.
\begin{align*}
&{\mathcal E}_{cf}-{\mathcal E}_{f}^w \\
=& \int_{\bm U} \int_{\bm C}  \big(p({\bm U})p({\bm C})- W_T({\bm U},{\bm C})p({\bm U},{\bm C})   \big)\\ 
&\cdot{\mathcal L}\big(f({\bm U},{\bm C}), y({\bm U}, {\bm C})\big)d{\bm U} d{\bm C} \\
\leq&\ \bigg|    \int_{\bm U} \int_{\bm C}  \big(p({\bm U})p({\bm C})- W_T({\bm U},{\bm C})p({\bm U},{\bm C})   \big)\\
&\cdot{\mathcal L}\big(f({\bm U},{\bm C}), y({\bm U}, {\bm C})\big) d{\bm U} d{\bm C}        \bigg|\\
\leq&\   sup_{g \in G}   \bigg|    \int_{\bm U} \int_{\bm C}  \big( p({\bm U})p({\bm C})- W_T({\bm U},{\bm C})p({\bm U},{\bm C})   \big)\\
&\cdot g( {\bm U},{\bm C}  )d{\bm U} d{\bm C}        \bigg|\\
=&  IPM_{G}\big(W_T({\bm U},{\bm C})p({\bm U},{\bm C}),p({\bm U})p({\bm C})\big)  .
\end{align*}
When $W_T({\bm U},{\bm C}) = \frac{p({\bm C})}{p({\bm C}|{\bm U})}$, we have:
$$IPM_{G}(W_T({\bm U},{\bm C})p({\bm U},{\bm C}), \ p({\bm U})p({\bm C})) = 0 ,$$
$${\mathcal E}_f^w = {\mathcal E}_{cf} .$$

\subsubsection{Proof of the causal prediction module.}

The optimal prediction probabilities of ad exposure are given by
\begin{equation*}
    {\bm c}^{rev*} = 
    {\rm arg}\ \mathop{{\rm max}}_{{{\bm c}^{rev}}} \sum_{k=1}^K \int_{{\bm v}^{rev}}  p({\bm v}^{rev}|{\bm c}^k)  {\rm log}\left({\bm c}_k^{rev} \right)   d{\bm v}^{rev} .
\end{equation*}
By maximizing value function and leveraging Lagrange multiplies, we can derive ${\bm c}^{rev*}$ by the following form
\begin{equation*}
    {\rm arg}\ \mathop{{\rm max}}_{{{\bm c}^{rev}}} \sum_{k=1}^K 
    \left(
     p({\bm v}^{rev}|{\bm c}^k)  {\rm log}\left({\bm c}_k^{rev} \right)
    \right) +
     {\chi}\left(\sum_{k=1}^K{\bm c}_k^{rev}-1 \right)  .
\end{equation*}
We have ${\bm c}^{rev}_k=-\frac{ p({\bm v}^{rev} | {\bm c}^{k}) }{\chi}=    \frac
    { p({\bm v}^{rev} | {\bm c}^{k}) }
    { \sum_{m=1}^K p({\bm v}^{rev} | {\bm c}^{m}) }$ by setting the above derivative to zero and solving ${\bm c}^{rev*}$.

Therefore, the objective $\mathop{{\rm min}}_{{{\bm v}^{rev}}} {\mathcal L}_{rev}$ for the learned representation ${\bm v}^{rev}$ becomes
\begin{equation*}
    \mathop{{\rm min}}_{{{\bm v}^{rev}}} \sum_{k=1}^K {\mathbb E}_{{\bm v}^{rev} \sim p({\bm v}^{rev} | {\bm c}_k)} 
    \left[{\rm log} 
    \frac
    { p({\bm v}^{rev} | {\bm c}^{k}) }
    { \sum_{m=1}^K p({\bm v}^{rev} | {\bm c}^{m}) }
    \right]  .
\end{equation*}
We can derive that
\begin{align*}
&\sum_{k=1}^K {\mathbb E}_{{\bm v}^{rev} \sim p({\bm v}^{rev} | {\bm c}_k) }
    \left[{\rm log} 
    \frac
    { p({\bm v}^{rev} | {\bm c}^{k}) }
    { \sum_{m=1}^K p({\bm v}^{rev} | {\bm c}^{m}) }
    \right] + K\ {\rm log}K \\
&= \sum_{k=1}^K 
    \left(
    {\mathbb E}_{{\bm v}^{rev} \sim p({\bm v}^{rev} | {\bm c}_k) }
    \left[{\rm log} 
    \frac
    { p({\bm v}^{rev} | {\bm c}^{k}) }
    { \sum_{m=1}^K p({\bm v}^{rev} | {\bm c}^{m}) }
    \right]
    + {\rm log}\ K
    \right) \\
&= \sum_{k=1}^K 
    {\mathbb E}_{{\bm v}^{rev} \sim p({\bm v}^{rev} | {\bm c}_k) }
    \left[{\rm log} 
    \frac
    { p({\bm v}^{rev} | {\bm c}^{k}) }
    {\frac{1}{K}\ \sum_{m=1}^K p({\bm v}^{rev} | {\bm c}^{m}) }
    \right] \\
&= \sum_{k=1}^K
    KL \left(
    { p({\bm v}^{rev} | {\bm c}^{k}) } \bigg|\bigg| {\frac{1}{K} \sum_{m=1}^K p({\bm v}^{rev} | {\bm c}^{m})}
    \right)\\
&= K \cdot JSD \left(p({\bm v}^{rev} | {\bm c}^{1}), ..., p({\bm v}^{rev} | {\bm c}^{K}) \right)  ,
\end{align*}
where $KL(\cdot||\cdot)$ is the KL divergence and $JSD(\cdot,...,\cdot)$ is the Jensen-Shannon Divergence \cite{Li18Proof,BicaAJS20} in the multi-distribution form. Because $JSD(\cdot,...,\cdot)$ is non-negative and equals zero when all distirbutions are equal and $K\ {\rm log}K$ is a constant, we have that $p({\bm v}_t^{rev}|{\bm c}_1)=...=p({\bm v}_t^{rev}|{\bm c}_K)$ by minimizing ${\mathcal{L}}_{rev}$.

\vspace{-3ex}
\subsection{Experiments}\label{appen:exp}

\subsubsection{Details of data preprocessing.} \label{appen:process}

The attribution modeling for bidding dataset published by \textbf{Criteo} company is widely deployed in the research of modeling user behavior and ad attribution\cite{Meynet_Criteo,RenFZLLZYW18,CAMTA}. As a sample of 30 days of Criteo live traffic data, this dataset has more than 16 million ad impression records and 45 thousand conversions over 700 ad campaigns. Each ad impression record contains items such as timestamp, user id, ad campaign, and side information. There is also a label denotes whether a click action has occurred, and the corresponding conversion ID if this sequence of ad impressions finally leads to a conversion. 
We preprocess the raw Criteo dataset in the following procedures: (i) we count the top 10 campaigns with the largest number of ad impression records and filter out the ad impression records corresponding to other campaigns; (ii) we group the ad impression entries, which have the same user id and conversion id, into the same sequence and sort each sequence by timestamp; (iii) for a conversion id of -1, i.e., for a specific user, a group of ad impression records that did not cause the user to convert, the original group is divided at a time interval of 3 days based on timestamp; (iv) we filter out ad sequences that are less than 3 in length; (v) we divide it into the train set and the test set, and ensure the set of user id in the test set is a subset of user id in the train set.

\subsubsection{Details of compared baselines.} \label{appen:baseline}

We compare \our with four kinds of baselines, \ie, statistical learning-based methods(SL), deep learning-based methods(DL), causal learning-based methods(CL) and ablations.

The statistical learning-based methods consist of three works:
\begin{itemize}[leftmargin=4mm]
    \item \textbf{LR}(Logistic Regression) model for ad attribution is proposed by Shao and Li\cite{ShaoL11}, in which channel's attribution values are calculated as the learned coefficients. 
    \item \textbf{SP}(Simple Probabilistic) model calculates the conversion rate taking into the conversion probability from the observed data into account. As in\cite{dalessandro2012causally}, the conversion rate is
    $$
        p(y=1|\{c_j\}_{j=1}^{m_i}) = 1-\prod_{j}^{m_i}(1-{\mathbf{Pr}}(y=1|c_j=k))  .
    $$
    \item \textbf{AH}(Additive Hazard) proposed by Zhang \etal \cite{zhang2014multi} is the first user conversion estimation model based on survival analysis and additive hazard function. 
\end{itemize}

The deep learning-based methods also consist of three works:
\begin{itemize}[leftmargin=4mm]
    \item \textbf{DNAMTA} is the Deep Neural Net with Attention Multi-touch Attribution model proposed by Arava \etal \cite{arava2018deep}. It leverages LSTM and attention mechanism to model the dynamic interaction between ad channels, and incorporates user-context information to reduce estimation bias.
    \item \textbf{DARNN} is the Dual-Attention Recurrent Neural Network proposed in \cite{RenFZLLZYW18} which uses dual-attention RNNs to combine both post-view and post-click attribution patterns together for the user conversion estimation. 
    \item \textbf{DeepMTA} is a phased-LSTM based model \cite{DeepMTA} which combines deep neural networks and additive feature explanation model for interpretable online multi-touch attribution. For fair comparison, we replace the phased-LSTM with vinilla LSTM.
\end{itemize}

The causal learning-based methods consist of two works:
\begin{itemize}[leftmargin=4mm]
    \item \textbf{JDMTA} is a causal-inspired model  \cite{du2019causally} which employs Shapley Value to compute the attribution credits for touchpoints. 
    \item \textbf{CAMTA} is the Causal Attention Model for Multi-touch Attribution proposed by Kumar \etal \cite{CAMTA}. This model leverages counterfactual recurrent network to minimize selection bias in channel assignment while conducting conversion estimation.
\end{itemize}

We also compare \our with its two ablations:
\begin{itemize}[leftmargin=4mm]
    \item  \ourrw removes the journey reweighting module and treats all journeys equally. It only employs the proposed causal conversion prediction model for MTA.
    \item \ourcasual replaces the causal RNN predictor with traditional RNN and only utilize the reweighting mechanism to eliminate the confounding bias of static user attributes.
\end{itemize}

\subsection{Generation of Synthetic Data}\label{appen:generation}

To generate synthetic data, we simulate the ad exposure policy and user conversions. 

\subsubsection{Details of ad exposure policy.} 
Ad delivery involves two issues: serving time and advertising channels. 
To generate the time series data for ad exposure, we can first use a Poisson process. Then, to assign the ad types for the events, we create a stochastic function.
In the dynamic-only setting, the intensity rate $\lambda_{exp}$ is a function of the user preference while the selection of ad types is independent of user preference. 
In the static-only setting, user characteristics are parameters of the stochastic function.
The user feature has an impact on both aspects of ad exposure in a hybrid setting.
We also simulate ad sequences whose generation is invariant across different user preferences to create unbiased data.

\subsubsection{Details of user conversion module.} The user conversion behaviour in multi-touch attribution can be viewed as occurrences in an inhomogeneous Poisson process\cite{tp_MTA_shender20}. 
In detail, we leverage a realization of a Poisson counting process combined with a time-varying intensity function, $\lambda(t)$. This process can be formulated as 
\begin{equation*}
    Y_i(t) - Y_i(s) \sim Poisson(\int_s^t\lambda(t)dt),
\end{equation*}
where $Y_i(t)$ is the number of conversion events for customer $i$ up until time $t$. 
As the user conversion is the effect of both ad exposures and user characteristics, we use a log-linear model for the intensity function, and allow it to depend on the impact of the previous seen ads and user features, then we have
\begin{equation*}
    \mathrm{log} (\lambda(t)) = \alpha_0 + \alpha_{user} + \sum_{j,k}g_k(t-t_j),
\end{equation*}
where $\alpha_0$ represents the log of conversion rate before any ads are shown regardless of user preferences, and $\alpha_{user}$ stands for the impact of user features. The item $g_k(t-t_j)$ models the impact of an ad exposure of channel $k$ that occurs at time $t_j$, which brings the jump in conversions and has an exponentially decaying effect.

\end{document}